*Original Article*

# Portuguese Households' Savings in Times of Pandemic: A Way to Better Resist the Escalating Inflation?

[1]**Ana Lúcia Luís**, [2]**Natália Teixeira**, [3]**Rui Braz**
[1]*Instituto Superior de Gestão, Portugal.*
[2]*Instituto Superior de Gestão, CEFAGE, Portugal.*
[3] *Instituto Português de Administração de Marketing, Portugal.*

***Abstract:*** *March 2020 confinement has shot Portuguese savings to historic levels, reaching 13.4% of gross disposable income in early 2021 (INE, 2023). To find similar savings figures we need to go back to 1999. With consumption reduced to a bare minimum, the Portuguese were forced to save. Households reduced spending more because of a lack of alternatives to consumption than for any other reason. The relationship between consumption, savings, and income has occupied an important role in economic thought [(Keynes, 1936; 1937); (Friedman, 1957)]. Traditionally, high levels of savings have been associated with benefits to the economy, since financing capacity is enhanced (Singh, 2010). However, the effects here can be twofold. On the one hand, it seems that Portugal faced the so-called Savings Paradox (Keynes, 1936). If consumers decide to save a considerable part of their income, there will be less demand for the goods produced. Lower demand will lead to lower supply, production, income, and, paradoxically, fewer savings. On the other hand, after having accumulated savings at the peak of the pandemic, the Portuguese are now using them to carry out postponed consumption and, hopefully, to better resist the escalating inflation.*

*This study aims to examine Portuguese households' savings evolution during the most critical period of the pandemic, between March 2020 and April 2022. The methodology analyses the correlation between savings, consumption, and GDP as well as GDP´s decomposition into its various components and concluded that these suddenly forced savings do not fit traditional economic theories of savings.*

## I. INTRODUCTION

Classical economic theory traditionally assigns savings a key role in the dynamics of the economy through its relationship with investment. A causal relationship is assumed between savings and investment based on the assumption that to have funds to invest in capital factor increases, it is necessary to give up consumption and make the funds available in the market. This relationship is documented in several studies which show that, in situations of economic stability, savings determine the ability of a country to finance itself for investment purposes, which will contribute to increasing productivity and the growth rate [(Feldstein and Horioka, 1980); (Baxter and Crucini, 1993); (Attanasio et al, 2000); (Holmes, 2005)]. The importance of savings is also notable in contexts of economic recession, as it helps countries cope better with the damaging effects of the crisis. From this perspective, savings is seen as a fundamental condition for economic growth, via investment, hence the relevance of its study. Along with the assumed importance of savings in economies, the reflection on households' savings and its impact on the macroeconomic context has been neglected as economies open to the exterior. As a result of globalisation, sources of finance and investment have diversified and expanded beyond the domestic market, relegating private savings to the background.

The savings function in an economy is determined by numerous economic, social, demographic, and cultural factors, with the determination of these factors essential for the correct determination of policies aimed at stimulating household savings. The relationship between savings, interest rates, and investment and the effects on productive investment and long-term economic growth potential is an old one, and there are different approaches to how this relationship occurs [(Keynes, 1936); (Kalecki, 1971); (Smith, 1776)]. Niculescu-Aron & Mihăescu (2012) identify the main determinants of population savings, highlighting that the level of development of an economy should be a fundamental parameter in the definition of actions to stimulate household savings, according to national specificities and household behavioural parameters. In more developed economies, aggregate saving increases in the face of economy-wide uncertainty, particularly when it affects labour income, implying that saving rates will continue to be maintained, or even increased, complicating the economic recovery process [(Cox et al., 2020); (Levenko, 2020); (Bayer et al., 2019); (Mody et al., 2012)].

Mody et al. (2012) validate the application of the precautionary savings model in periods of economic crisis and in face of uncertainty, especially of labour income, with increased savings rates contributing to lower consumption and GDP growth. In household saving behaviour, whether due to individual motives (household characteristics) or existing liquidity constraints (institutional macroeconomic variables), there is evidence of some homogeneity concerning saving preferences and the relative importance of different motives to save (Le Blanc et al., 2014). With the increased uncertainty resulting from such a scenario, financial crises tend to increase savings rates, leading to a contraction in consumption and GDP growth. The concept of precautionary savings in the face of uncertainty generated the economic crisis, generating greater uncertainty of labour income, leading to increased household savings [(Mody et al., 2012); (Steindl, 1990)].

During the Covid-19 pandemic, savings rates among European households reached record levels, and there are different possible reasons: precautionary motivations induced by increased economic uncertainty, reduced consumption due to confinements, and the expected increase in the tax burden on households soon due to increased public debt (Gropp & McShane, 2021). Gersovitz (1988) evaluated savings from the perspective of decision-makers opting for current and future consumption. Thus, it can be established that savings tend to increase in periods of economic uncertainty, especially in periods of economic crises. With the crisis generated by the COVID Pandemic, it is important to check whether the paradox of thrift holds. Keynes (1936) states that the aggregate impact of an increase in the propensity to save by households leads to a reduction in the economy's output, further accentuating the recessionary impact. It is natural for individuals and households to postpone consumption and expenditure in the face of economic recession, but it is also easy to anticipate that the reduction in collective expenditure will lead to a fall in aggregate output, via consumption. According to Al-Wakil (2020), the collective fear and unpredictability of what will happen fuels economic recession. However, the nature of the recession generated by the COVID pandemic has led to savings being increased by the forced closure of economic activity, not because of the fear of unemployment, but the inability to spend (Al-Wakil, 2020).

Portugal's traditionally low savings rate has been an undervalued fact. Despite the acknowledgement of the issue, little has been done to address the problem. Hence, the recent sudden increase, to historical values, of the savings rate in Portugal, caused by the pandemic and the confinements, deserves special attention. Many factors may be pointed out as causing the low savings rate in Portugal. Whether it is a direct consequence of the Portuguese financial development model, the social security distribution system, or the variables typically described in economic literature such as economic growth, the unemployment rate, demographic trends, interest rates or the level of inflation, what is certain is that the savings figures in Portugal are systematically below most of the European Union (EU) countries.

**Figure 1 - Pre-Pandemic Savings**

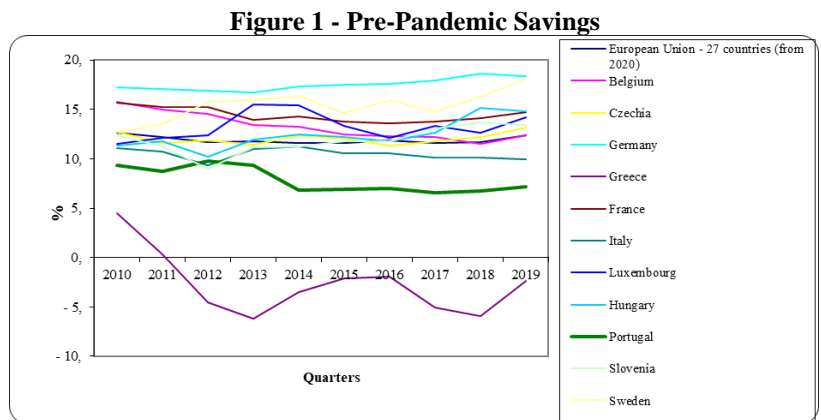

*Source: Pordata (2023)*

Figure 1 presents a selection of European countries, all of which have savings rates above the Portuguese rate. Portugal's weak position in the last decade is evident, clearly below the EU-27 average. Only Greece, Lithuania, Cyprus, and Poland (the last three are not shown in the graph) have savings rates lower than those of Portugal.

*A) Savings and the Pandemic Crisis*

In 2020 and 2021 we witnessed what is conventionally called the macroeconomics of the pandemic (Pires de Souza, 2021), with two striking features. On the one hand, the interdiction of consumption and production through public health policies and, on the other hand, an important process of income redistribution, a direct consequence of economic policy measures taken by governments to minimize the effects of the paralysis of economic activity. The combination of these two characteristics increased household savings. The macroeconomics of the pandemic is a macroeconomics of recession. However, this was not a recession like all the others studied by economists since the great depression of the 1930s. In the face of epidemics like COVID-19, social isolation was the main instrument for saving lives. And this isolation avoided the agglomerations that normally occur in the consumption of goods and services and face-to-face work, the so-called contact-intensive activities. In other words, social isolation required the reduction of consumption and production and, therefore, recession. The recession, therefore, worked as a public health measure (Baldwin and di Mauro, 2020).

It is important to analyse the effects of crises on savings behaviour. In recessionary contexts, a counter-cyclical effect is visible in the evolution of household savings, including in the Portuguese economy. Several studies document the situation in different countries, where there is an increase in household savings, despite the recession, mainly stimulated by the precautionary motive [(Kim, 2010); (Nahmias, 2010)].

Adema and Pozzi (2015) studied household savings in 16 OECD countries over the period 1969-2012 and suggest that in periods of economic recession household uncertainty about labour income grows, thus triggering higher precautionary savings. In recessionary situations, it is also usual to place causality in the opposite direction. In this context, if there is a generalised increase in household savings, there is evidence that the economy falls into the so-called Savings Paradox, a concept developed by Keynes in 1936. The concept states that if all economic agents try to increase savings at the same time, such behaviour will cause a sharp drop in aggregate demand, via consumption, which in turn will have negative effects on the growth of the economy and will end up minimising the initial effort to save (Mota, 2017).

Several studies have been developed regarding households' saving behaviour during the COVID-19 pandemic. Cox et al. (2020) determined that all households, regardless of income, cut spending at the start of the pandemic, with high-income households cutting spending more than low-income households but with small differences. In terms of recovery, the differences are substantial, with expenditure recovering much more quickly for lower incomes (Bachas et al., 2020). Vanlaer et al. (2020) link consumer confidence and household saving behaviours in Europe, determining which specific indicators of consumer sentiment played the most significant role, in particular confidence in household financial situation. Dossche & Zlatanos (2020), determine that in the euro area, in response to COVID-19, the household savings rate reached unprecedented levels in the first half of 2020. The containment measures followed by most European countries that prohibited households from consuming a large part of their normal expenditure basket led to forced savings; but also, the sudden outbreak of the pandemic made the risk of future unemployment skyrocket, leading to precautionary savings, this being the main driver of the increase in household savings (Dossche & Zlatanos, 2020).

Brewer & Patrick (2021) analyse that the overall decrease in spending during the pandemic, driven by decreases in leisure activities, meals out, holidays, and commuting costs (and a commensurate increase in savings) does not reflect the experience of all households, emphasising the importance of properly understanding differential experiences and the unequal impacts they have on the living standards of households with different incomes. Martin et al. (2020) developed a microeconomic model to estimate the direct impact of remoteness on household income, savings, consumption, and poverty levels, assuming a crisis period where part of households has income drops (using accumulated savings to meet consumption); and a recovery period, where households try to restore savings to the pre-crisis level. In consumption patterns, household consumption was radically altered during this period with an initial sharp increase (particularly retail and food), followed by a drastic fall in overall spending (Baker et al., 2020). Andersen et al. (2022) suggest that the contraction in consumption during confinement was caused by temporary health risks and product sourcing constraints, with existing savings being little related to persistent negative spillovers. On the other hand, Basselier & Minne (2021) analysed the behaviour of savings that will tend to be permanent after the COVID-19 crisis, with only a very limited share of these excess savings returning to consumption.

## II. METHODS

From the literature review carried out, several hypotheses can be presented and assessed in the context of Portuguese reality:

H1: In periods of economic crisis (such as the 2008 financial crisis and the pandemic crisis), Portuguese households' savings have increased due to uncertainty and prevention.
H2: Confidence in the behaviour of the economy, in particular in the financial situation, has a positive impact on savings.
H3: The increase in savings that occurred during the COVID period was driven by an exogenous factor and had sustained behavioural economic consequences.

To evaluate and test the hypotheses in the context of Portugal, an evolutionary study of the savings rate of Portuguese households was conducted, focusing on the COVID pandemic and the periods of imposed confinement. Simultaneously, correlations were calculated between the level of savings, consumption, and the evolution of GDP in Portugal, and a breakdown of GDP into its different components was carried out, to assess the veracity of the aforementioned hypotheses as well as whether the pattern of savings, particularly when imposed in a sudden and unforeseen manner, fits in with traditional economic theories of savings. During the period under analysis, various sources were used as references in the collection of statistical data to study the behaviour of Portuguese households concerning savings and in comparison to the behaviour of economic agents in Europe, such as the Bank of Portugal (BdP), Instituto Nacional de Estatística (INE), Pordata and Eurostat.

### III. RESULTS AND DISCUSSION

In 2020 Gross Domestic Product (GDP) was €200.1 billion, which represented a nominal decrease of 6.7%. There had been an increase of 4.5% in 2019. With this result, 2020 was the year with the largest contraction in economic activity since 1995 (the year in which a series of INE's National Accounts publications began). The Gross Disposable Income (GDI) of the economy decreased by 5.5%, having registered a positive change of 4.4% in the previous one. The nominal Gross Disposable Income of households (including households and Non-Profit Institutions Serving Households-NPISHs) reached €146.8 billion in 2020, which represented a decrease of 0.7% relative to the previous year, compared to an increase of 4.6% seen in 2019. Household Final Consumption expenditure decreased, as it turned out, sharply. It recorded a drop of 6.4% in 2020 while in 2019 it had seen an increase of 4.1%. Due to the reduction seen in consumption, much greater than the reduction seen in GDI, Household savings registered a considerable increase, of more than 76.5%, which corresponds to a rate of 12.8% as a percentage of GDI. In 2019, this savings rate was 7.2% (see Table 1).

**Table 1 - Annual Indicators for the Portuguese Economy**

|  | 2017 | 2018 | 2019 | 2020 | 2021 |
|---|---|---|---|---|---|
| **Nominal GDP (rate of change)** | 5,1 | 4,7 | 4,5 | -6,7 | 5,6 |
| **Gross Disposable Income (rate of change)** | 5,6 | 4,5 | 4,4 | -5,5 | 6,7 |
| **Households and NPISHs saving rate (% of GDI)** | 6,6 | 6,8 | 7,2 | 12,8 | 10,9 |
| **Households' Final Consumption and NPISHs (rate of change)** | 3,7 | 4,2 | 4,1 | -6,4 | 5,8 |
| **Households Disposable Income and NPISHs (rate of change)** | 3,1 | 4,3 | 4,6 | -0,7 | 4,0 |

*Source: INE (2023)*

Disaggregating the data by quarters, it is evident the increase in the savings rate (gross saving as a percentage of disposable income, gross) from the first to the second quarter of 2020, where there was an increase from 7.8% to 10.9%. From 2020 to 2021 Portuguese households continued to save. It is observed that the peak of the savings rate, with 13.4% of disposable income, occurred in the first quarter of 2021, corresponding to the second confinement in Portugal. To find such high values in Portugal, it is necessary to go back more than twenty years, to the fourth quarter of 1999, with a savings rate of 13.7%. The increase in the first quarter of 2021 was influenced, once again, by the reduction in consumption expenditure that managed to compensate for the slight decrease in disposable income.

**Figure 2 – Saving Rates in Portugal**

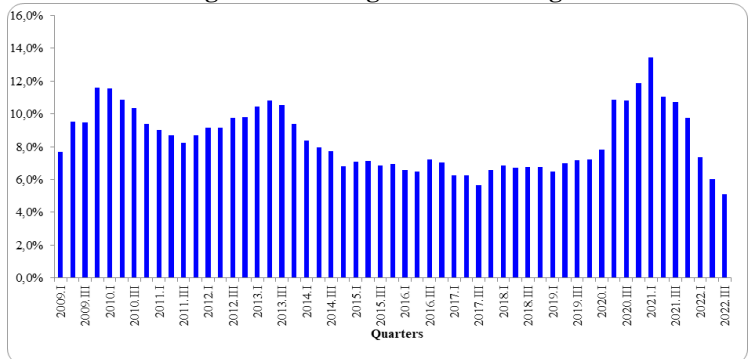

*Source: INE (2023)*

Despite the serious economic and social crisis that the confinements have caused in Portugal, disposable income has remained relatively stable, according to data from INE. Since March 2020, the Government, while imposing many businesses to close and reduce their activity and millions of workers to stay at home, advanced with a variety of income support and job retention. The simplified layoff was the strongest of these measures, allowing to maintain jobs and save a good part of the salaries. Unemployment rose, but not as dramatically as in previous crises. The moratorium on credit agreements allowed Portuguese families to reduce their expenditure on housing. Thus, with disposable income stabilised, many families simply adjusted their spending habits. Consumption fell sharply and what was not spent on cafés, restaurants, transport, clothing and footwear, gyms, and housing, among others, was almost entirely translated into more savings.

No consumer can claim that they have not felt any impact of the pandemic, even though it has had differentiated impacts. It is a classic effect that, in times of crisis, higher-income households manage to build up a more generous savings pillow than lower-income households. According to Banco de Portugal (2021: 11), "the evidence points to savings accumulation in 2020 has been more concentrated in higher income households, whose marginal propensity to consume is lower. Bank card purchases show that high-consumption individuals reduced their spending more sharply in 2020 than low-consumption individuals." Evidence shows that income losses, whether due to layoffs or unemployment, have occurred mainly in groups of workers with lower wages, fewer qualifications, and precarious labour contracts. This is typically the universe where, already under normal conditions, there is a low saving capacity, given that the day-to-day expenses absorb a large part of the income.

**Figure 3 – Contribution of the various components to Portuguese GDP variations**

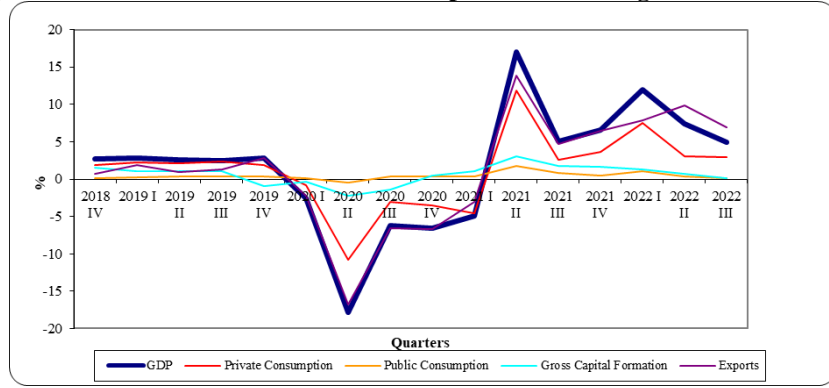

*Source: INE (2023)*

Note: GDP at market prices, in volume, percentage points, adjusted for calendar and seasonal effects.

When breaking down the variation in GDP by quarters into its various components, it is clear, as expected, the weight of private consumption. While in 2019 the weight of consumption in the variation of GDP never exceeded 2,3 p.p., in 2020 and 2021 this value was responsible, together with exports, for the variations in GDP, which demonstrates the effect of the constraints on consumption. Thus, during the first confinement in Portugal, starting in March 2020 and corresponding to the second quarter, consumption contributed with a negative variation of 10.8 p.p. and exports with a decrease of 16.8 p.p. Together with an increase in imports of 12.5 p.p., they combined for a drop in GDP of 17.8 p.p., the largest fall in GDP in Portugal during the pandemic crisis. In the first quarter of 2021, corresponding to the second confinement, the drop in GDP was much less pronounced, falling by 4.9 p.p. Consumption decreased by 4.6 p.p. and exports by 3.1 p.p. Of note is the increase in GDP in the following quarter, which corresponds to the easing of the restrictive measures imposed on the economy. In the second quarter of 2021, GDP increased by 17 p.p., with private consumption accounting for 11.9 p.p. of this increase and exports by 13.8 p.p..

**Figure 4 – Contribution to the Portuguese Household Saving Rate**

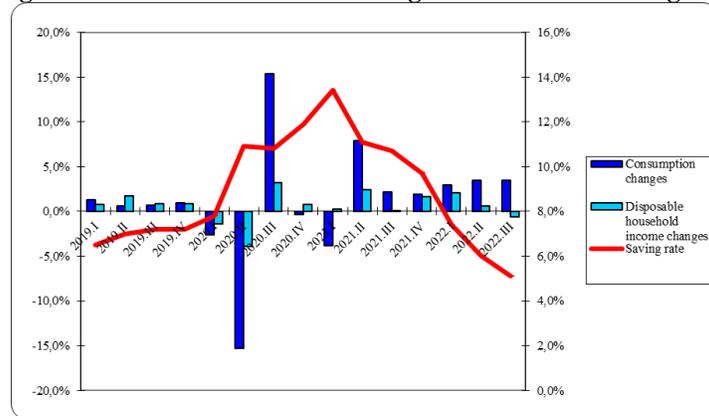

*Source: INE (2023)*

In the first quarter of 2020, the date of the beginning of the first confinement in Portugal, the savings rate was 7.8% of disposable income. This value does not yet reflect the real dimension of the restrictive measures, since the confinement started in mid-March. In the second half of the year, with confinement already in full effect, we witnessed an increase in the savings rate to 10.9% of disposable income. This result was a consequence of the very sharp drop in private expenditure, which fell 15.3% compared to the previous quarter and which more than offset the drop in disposable income by 3.9%. In the third quarter of 2020, with the gradual deconfinement of the economy, private consumption grew by 15.4% compared to the previous quarter. Disposable income registered a change of 3.2% and the savings rate was 10.8%. In the last quarter of 2020, with the deterioration of the pandemic situation in Portugal but still, with no new confinements, private consumption decreased by only 0.3% compared to the previous quarter, and disposable income registered an increase of 3.2%. Savings, reflecting the uncertainty and insecurity presented by the new wave of the pandemic, have risen to 11.9% of disposable income.

In the first quarter of 2021, the country enters the second confinement. As a result, the household savings rate rose again, reaching 13.4% of disposable income, which corresponded to an extremely high value, given the traditionally low values of savings in Portugal. This result was a consequence of the 3.9% reduction in final consumption expenditure, compared to the previous quarter, which exceeded the slight 0.2% increase in disposable income. In the second quarter of 2021, with the gradual resumption of economic activities, although with some restrictions, private consumption increases again by 7.9% compared to the previous quarter, disposable income increases by 2.4% and the savings rate is 11.1%. Subsequently, and with the gradual deconfinement of the Portuguese economy, the savings rate shows a decreasing pattern, being already in the pre-pandemic values in the first two quarters of 2022, with 7.4% and 6% of disposable income, respectively. This change in savings pattern can be justified by deconfinement but also by the need to cope with the price increases that resulted from the simultaneous reopening of the global economy.

As expected, and predicted, the savings levels witnessed during the confinements quickly volatilised. In the period from September 2021 to September 2022, the savings rate of Portuguese families fell to values close to zero. From 7.75% gross savings rate in September 2021 to 0.24% in September 2022. According to Eurostat, in the third quarter of 2022, Portugal was the country with the lowest savings rate among the Eurozone countries, which had a value of 13.2%. Expenditure on postponed consumption, the rise in interest rates and the escalating inflation rate are at the root of the sharp fall in savings in Portugal. In 2022, the average inflation rate in Portugal was 7.8%, the highest since 1992, a value that certainly contributed to the eroding of the savings rate. The forecasts of the Bank of Portugal (2021) for the savings rate to return to the pre-pandemic level in the period 2021-2023 did not materialise, with a sharp drop to historic lows already seen in 2022. The drop in the savings rate can be justified by the opening of the economy after a period of one-off and imposed closure which was associated with an inflationary crisis, resulting from blockages in the global distribution chain and the war in Ukraine, which is compounded by the average income level in Portugal.

**Figure 5 - Post-Pandemic Gross Savings Rate**

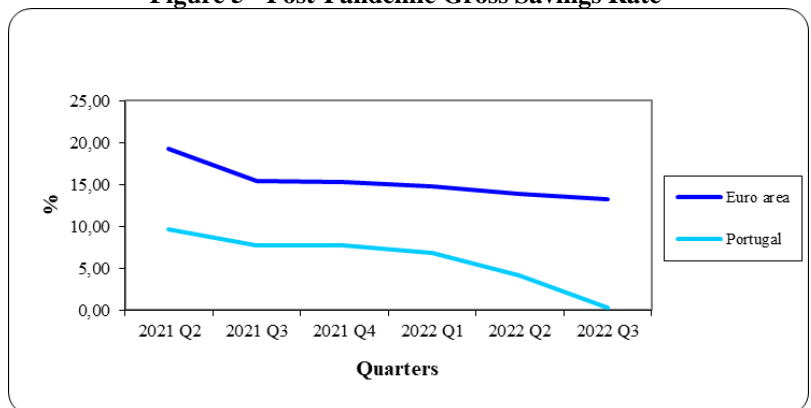

*Source: Eurostat (2023)*

The theoretical approach presented justifies the treatment of Gross Disposable Household Income (GDHI), GDPpc and Consumption (C) as explanatory variables of aspects of Saving (S) behaviour as the dependent variable. Through the correlation analysis, we conclude that all coefficients are negative, ranging from -0.23 for the correlation where GDHI is the independent variable to -0.62, corresponding to the correlation where C is the independent variable. Thus, the reduction in GDHI, in GDPpc and Consumption increases the tendency to save. The values of the correlations reinforce the literature which indicates that uncertainty and fear lead to increased savings for prevention reasons.

**Table 2 - Correlations**

|       | GDHI  | GDPpc |
|-------|-------|-------|
| GDHI  |       |       |
| GDPpc | 0,96  |       |
| C     | 0,88  | 0,97  |
| S     | -0,23 | -0,43 |

*Source: Author´s calculations; INE (2023)*

This study could be extended to analyse, in a specific way and using causality tests, the causal relationships established between the growth rates of savings, investment and GDP.

## IV. CONCLUSION

This research aims to examine Portuguese households' savings evolution during the most critical period of the pandemic. The study is elaborated with data regarding the Portuguese economy, being made with occasional comparisons with the European average and some specific countries. The theory holds that savings are made in economic recession scenarios as an essentially preventive way and as an anticipation of potential future difficulties. In the case of the economic crisis resulting from the COVID pandemic, some additional factors have to be added: the closure of the economy imposed by the government and forced saving due to the absence of where to spend. Thus, the precautionary motive is replaced by an 'absence of alternative' motive. The economic awareness of the need to save (either out of fear or because one intends to consume in the distant future) is replaced.

Regarding the hypotheses put forward, the following conclusions can be drawn:

H1: In periods of economic crisis (such as the 2008 financial crisis and the pandemic crisis), Portuguese households' savings have increased due to uncertainty and prevention. This is true even though it is fundamental to bear in mind the unique characteristics of the economic crisis resulting from the pandemic, where the compulsive closure of the economy has strongly conditioned the capacity and decision of families to consume.

H2: Confidence in the behaviour of the economy, in particular in the financial situation, has a positive impact on savings. Not verifiable. The data analysed point to a decrease in savings after the economic crisis, being in values practically null, contradicting the literature. It is important to analyze the impact of the rising inflation in the post-pandemic period which may have a relevant impact on this analysis. In a future analysis and when relevant statistical data is available, this issue can be further analysed.

H3: The increase in savings that occurred during the COVID period was driven by an exogenous factor and had sustained behavioural economic consequences. This hypothesis is not true. It is true Saving has indeed been due to an exogenous factor since it was "imposed" by the compulsory closure of the economy. However, the current data on the level of savings, which are at historical values and almost zero, makes the behavioural change that was expected to occur impossible.

In future work, a segmented analysis of saving by income level could be developed, to determine to what extent and how households' saving behaviour is conditioned by disposable income.

**Interest Conflicts**
The author(s) declare(s) that there is no conflict of interest concerning the publishing of this paper."